\documentclass[amsmath,amssymb, reprint]{revtex4-1}

\usepackage{graphicx}
\usepackage{dcolumn}
\usepackage{bm}

\usepackage[utf8]{inputenc}
\usepackage[T1]{fontenc}
\usepackage{mathptmx}
\usepackage{etoolbox}

\begin{document}

\title[Elastic properties of Yukawa crystals]{Elastic properties of Yukawa crystals}

\author{A.~A. Kozhberov}
\email{kozhberov@gmail.com}
\affiliation{Ioffe Institute, Politekhnicheskaya 26, 194021, Saint Petersburg, Russia}%

\date{\today}

\begin{abstract}
We study elastic properties of solid Yukawa systems. Elastic moduli and effective shear modulus of body-centered cubic (bcc) and face-centered cubic (fcc) lattices are obtained from electrostatic energies of deformed crystals. For the bcc lattice our results are well consistent with previous calculations and improve them, while results for the fcc lattice are mostly new. We have also obtained an analytical expression of the elastic moduli for the weak polarization and constructed a convenient approximation for the higher polarization.  
\end{abstract}

\maketitle

\section{Introduction}
A model of a plasma composed of the point-like particles interacting via an effective Yukawa potential is quite regularly applied to complex (dusty) plasmas. The neutralizing background in this system is non-degenerate, while the typical charge number of interacting particles is about $10^3$. At sufficiently low temperatures, such Yukawa system solidifies with formation either body-centered cubic (bcc) or face-centered cubic (fcc) crystal (e.g., Refs.~\onlinecite{HFD97,Fort1,Fort2}, and references therein). 

A similar model is used for description matter in solid interiors of the white dwarfs and neutron stars (e.g., Refs.~\onlinecite{B02,B15}). The background in this model consists of degenerate electrons, while point-like ions interact via screened Coulomb potential with the polarization corrections in the small-wavenumber limit of the linear response theory, which is equivalent to the Thomas-Fermi approximation in the electron-ion Coulomb plasma model (see, e.g., Ref.~\onlinecite{HPY07}, and references therein). 

In the current paper, we assume that particles, which form the lattice, are fixed in its nodes and do not move. So thermal and quantum fluctuations are neglected and we restrict ourselves to considering only the electrostatic energy. In turn, this electrostatic energy can be described by the same equation for both systems\cite{K18}, and therefore, in this approximation, both systems also have the same elastic constants (they could be obtained from electrostatic energies of deformed lattices\cite{F36,K19,C21}). In Refs.~\onlinecite{K19,C21} the elastic properties of solids with the uniform background were considered. Here we extend our method to a detailed study of crystals with an arbitrary screening parameter.

As sound velocities along high symmetry directions are related to elastic constants, our calculations can be important for studying various waves and their propagation in dusty and colloidal plasma, which have been observed in different experiments (e.g., Refs.~\onlinecite{BM95,FK00,T09,Y04,MP10,RSW15,YK19}) and interest in such researches continues unabated. However, from a theoretical point of view, not all questions have yet been resolved.
For the first time from molecular dynamics simulations the elastic properties of dusty Yukawa crystals were considered in Ref.~\onlinecite{RKG88}, where elastic moduli of bcc and fcc lattices were obtained. 
Further, in Ref.~\onlinecite{B15}, the first order polarization corrections to the elastic moduli of uniform Coulomb crystals in degenerated matter were calculated for the bcc lattice via perturbation theory. In Ref.~\onlinecite{Kh19} sound velocities of different dusty Yukawa systems in arbitrary screened regime were calculated analytically. A distinctive feature of Ref.~\onlinecite{Kh19} is that it takes into account the background contribution to the internal energy,\cite{K14,KK20} while in our work, it is not done. For quantities, which do not depend on the elementary cell volume change, this difference is not relevant. 

For the degenerated stars, which are the main point of our interest, the knowledge about elastic properties is necessary for various high-precision starquake and oscillations researches (e.g., Refs.~\onlinecite{D98,U00,G18}). For instance, in Ref.~\onlinecite{S91} it was demonstrated that toroidal, spheroidal and interfacial oscillation modes significantly depend on the effective shear modulus. Which in turn used for interpretation of some observational data (e.g., Refs.~\onlinecite{SW06,HP17}). In Refs.~\onlinecite{BL14,LBL16} it was shown that the magnetar activity is generated by the shear motions in the neutron star crust and therefore depends on its elastic properties. Also it is not excluded that the next generation gravitational wave detectors will discover gravitational waves from irregularities on the surface of neutron stars, in particular mountains (e.g., Refs.~\onlinecite{H06,JO13,HP17}).

Consequently, the studies of the elastic properties of Yukawa crystals presented in the current paper have not lost their relevance and could be used in various branches of physics.

\section{Electrostatic energy}
\label{ele}
The model of a Yukawa crystal assumes that the system consists of point-like particles with the charge number $Z$ and the number density $n$ on a neutralizing background, which are arranged into a lattice and interact via the Debye-screened Coulomb potentials with the screening wavenumber $\kappa_{\rm D}$. 
As shown in Ref.~\onlinecite{HFD97}, in the ground state these particles could form either bcc or fcc lattice type. We represent both lattices as simple cubic lattices with main translation vectors: $\textbf{a}_1=a_{\rm l}(1,0,0)$, $\textbf{a}_2=a_{\rm l}(0,1,0)$,  $\textbf{a}_3=a_{\rm l}(0,0,1)$, where $a_{\rm l}$ is the lattice constant, and with $N_{\rm cell}$ basis vectors ($N_{\rm cell}=2$ for the bcc lattice and $N_{\rm cell}=4$ for the fcc lattice). This definition allows us to consider deformations of both lattices simultaneously.

In Coulomb crystals, which are used in the theory of degenerate stars, the point-like particles are ions, while the background is formed by relativistic degenerate electrons. Background is not uniform and can be described by the dielectric function with the Thomas-Fermi electron screening parameter $\kappa_{\rm TF}$.
In dimensionless form it could be written\cite{HPY07}
\begin{equation}
\kappa_{\rm TF}a\approx 0.1850\,Z^{1/3}\,\frac{(1+x_{\rm r}^2)^{1/4}}{x_{\rm r}^{1/2}}~,
\label{kTFa}   
\end{equation}
where $a \equiv (4\pi n/3)^{-1/3}$ is the Wigner-Seitz radius,
\begin{equation}
x_{\rm r}\equiv \frac{\hbar(3 \pi^2 Z n)^{1/3}}{m_e c} \approx 0.01\left(\rho \frac{Z}{M}\right)^{1/3}~ \label{x}   
\end{equation}
is the electron relativity parameter, $m_e$ is the mass of electron, $M$ is the mass number of point-like particles, $\rho$ is their mass density in g~cm$^{-3}$. 

The equation for the electrostatic energy of Coulomb crystal with a polarized background was obtained in Ref.~\onlinecite{B02}:
\begin{eqnarray}
U_{\rm M} &=& NZ^2e^{2}
\nonumber \\
&\times&\left[\frac{1}{N_{\rm cell}}\sum\limits_{l,p,p'} \left(1-\delta_{l0}\delta_{pp'}\right) \frac{E_{-}+E_{+}}{4Y_{lpp'}}\right. \nonumber \\
&+&\frac{2\pi n}{N_{\rm cell}^2}\sum\limits_{m,p,p'} \frac{{\rm exp}\left(-\frac{G_m^2+\kappa^2}{4 A^2}-i{\bf G}_m({\bm \chi}_p-{\bm \chi}_{p'})\right)}{G_m^2+\kappa^2}\nonumber \\
&-&\left.\frac{\kappa}{2}{\rm erf}\left(\frac{\kappa}{2  A}\right)-\frac{ A}{\sqrt{\pi}}{\rm exp}\left({-\frac{\kappa^2}{4 A^2}}\right)-\frac{2\pi n}{\kappa^2} \right]~, \label{UTF}
\end{eqnarray}
where $l=\{l_1, l_2, l_3\}$ and $m=\{m_1, m_2, m_3\}$ are arbitrary integers, sums over $p$ and $p'$ go over all point-like particles in the elementary cell ($p,p'=1 \dots N_{\rm cell}$), 
${\bf Y}_{lpp'}={\textbf{R}}_l+{\bm \chi}_p-{\bm \chi}_{p'}$, ${\bm \chi}_p$ are the basis vectors, $E_{\pm}=e^{\pm\kappa Y_{lpp'}}\left[1-{\rm erf}\left({ A}Y_{lpp'}\pm \kappa/(2 { A})\right)\right]$ and ${\rm erf}(z)$ is error function, $ A$ is chosen so that summation over direct (${\bf R}_l=l_1 \textbf{a}_1+l_2 \textbf{a}_2+l_3 \textbf{a}_3$) and reciprocal (${\bf G}_m=m_1 \textbf{g}_1+m_2 \textbf{g}_2+m_3 \textbf{g}_3$) lattice vectors converge equally rapidly, usually ${ A} a\approx2$. 

In Eq. (\ref{UTF}) the subscript in $\kappa$ is omitted because, as shown in Ref.~\onlinecite{K19}, $U_{\rm M}$ has the same form for $\kappa_{\rm TF}$ and $\kappa_{\rm D}$. Note that usually Eq. (\ref{UTF}) is used for arbitrary $\kappa_{\rm D} a$ (e.g., Ref.~\onlinecite{HFD97}), but only for $\kappa_{\rm TF} a \ll 1$ (e.g., Ref.~\onlinecite{B15}, see Ref.~\onlinecite{K18} for details).

At $\kappa a \ll 1$ the electrostatic energy can be written as 
\begin{equation}
U_{\rm M}=N\frac{Z^2e^2}{a}\left(\zeta+\eta (\kappa a)^2\right)~, 
\label{U2TF}
\end{equation}
where $\zeta$ is called a Madelung constant, and both $\zeta$ and $\eta$ depend only on the type of the lattice. For the bcc lattice:
\begin{eqnarray}
\zeta_{\rm bcc}&=&-0.895929255682~,\\
\eta_{\rm bcc}&=&-0.103732333707~,
\end{eqnarray}
while for the fcc lattice: 
\begin{eqnarray}
\zeta_{\rm fcc}&=&-0.895873615195~,\\
\eta_{\rm fcc}&=&-0.103795687531~.
\end{eqnarray}
There is no term proportional to $\kappa a$ in Eq. (\ref{U2TF}), since at $\kappa a=0$ there should be an extremum. Also the Thomas-Fermi dielectric function depends on $\kappa^2$.
For higher $\kappa a$, beyond the quadratic approximation Eq. (\ref{U2TF}), $U_{\rm M}$ was explored in Ref.~\onlinecite{KP21}.

In Eq. (\ref{U2TF}) we use ordinary units $NZ^2e^2/a$, while $\zeta$ and $\eta$ are normalized to the Wigner-Seitz radius $a$. However, in some cases volume of the elementary lattice changes and it is more convenient to use a $2a_{\rm l}$ lattice constant units:
\begin{equation}
\tilde{\zeta}\equiv 2 \zeta \frac{a_{\rm l}}{a}~, \qquad
\tilde{\eta}\equiv 2 \eta \frac{a_{\rm l}}{a}~.
\end{equation}
By the same principle, pressure and elastic coefficients are determined in the next section.

\section{Elastic properties}

Eq. (\ref{UTF}) can be applied for any set of main translation vectors.
Therefore, we use it for different deformed lattices.

The simplest case of deformation is the stretching along edges of the main lattice cube. 
Under the influence of this deformation vector $a_{\rm l}(l_1,l_2,l_3)$ translates as
\begin{equation}
a_{\rm l}(l_1,l_2,l_3)\to a_{\rm l}\left(l_1,c_1 l_2, c_2 l_3,\right)~,
\label{e:trans1}
\end{equation}
where $c_1$ and $c_2$ characterize the stretch value, while $a_{\rm l}$ is constant. 
In this case the electrostatic energy $U_{\rm M}$ becomes a function of $c_1$ and $c_2$.
If the lattice deformations are small, we can expand $U_{\rm M}(c_1,c_2)$ in powers of $c_1-1$ and $c_2-1$. 

For a uniform background it all boils down to the Madelung constant changing
\begin{eqnarray}
\tilde{\zeta}(c_1,c_2) &\approx& \tilde{\zeta}-\frac{1}{3}\tilde{\zeta}\left[(c_1-1)+(c_2-1)\right]\nonumber \\
&+&\frac{1}{4} \tilde{s}^{xxxx}\left[(c_1-1)^2+(c_2-1)^2\right] \nonumber \\
&+&\frac{1}{2}\tilde{s}^{xxyy}(c_1-1)(c_2-1)~,
\label{SM2}
\end{eqnarray}
where
\begin{eqnarray}
\tilde{s}^{xxxx}_{\rm bcc}&=&-1.48480792~, \\
\tilde{s}^{xxyy}_{\rm bcc}&=&-0.47067387~, \\
\tilde{s}^{xxxx}_{\rm fcc}&=&-1.89664945~, \\
\tilde{s}^{xxyy}_{\rm fcc}&=&-0.57996263~.
\end{eqnarray}
These coefficients were obtained earlier in Refs.~\onlinecite{F36,OI90,II03,B11,B15,K19} and are fully consistent with ours.

From Eq. (\ref{SM2}) it follows that for $\kappa a=0$ the electrostatic pressure is
\begin{equation}
P_0=n\frac{Z^2e^2}{a} \frac{\zeta}{3}~,
\end{equation}
and elastic coefficients (see Refs.~\onlinecite{B11,B15}) are 
\begin{equation}
c_{11} \equiv n\frac{Z^2e^2}{a}s^{xxxx}~, \quad c_{12} \equiv n\frac{Z^2e^2}{a}\left(s^{xxyy}+\frac{\zeta}{3}\right)~.
\end{equation}

Since the stretching changes the volume of the elementary cell, it also changes $\kappa a$. So if we want to calculate the first order polarization corrections to $P_0$, $c_{11}$ and $c_{12}$ at $\kappa a \ll 1$, we have to expand $\zeta$, $\eta$ and $\kappa a$ in powers of $c_1-1$ and $c_2-1$. 

For a degenerate electron background, see Eq. (\ref{kTFa}), this gives 
\begin{eqnarray}
P^{\rm scr}&=&n \frac{Z^2e^2}{a}(\kappa a)^2 \frac{\eta}{3}\frac{x_{\rm r}^2}{1+x_{\rm r}^2} \label{Pscr}~, \\
c_{11}^{\rm scr}&=& n\frac{Z^2e^2}{a} (\kappa a)^2 \nonumber \\ &\times&\left[2\chi+\frac{2\eta}{9}+\frac{5\eta}{9}\frac{x_{\rm r}^2}{1+x_{\rm r}^2}-\frac{\eta}{9}\left(\frac{x_{\rm r}^2}{1+x_{\rm r}^2}\right)^2\right]~, \label{c11scr} \\
c_{12}^{\rm scr}&=& n\frac{Z^2e^2}{a} (\kappa a)^2 \nonumber \\ &\times&\left[-\chi-\frac{\eta}{9}+\frac{5\eta}{9}\frac{x_{\rm r}^2}{1+x_{\rm r}^2}-\frac{\eta}{9}\left(\frac{x_{\rm r}^2}{1+x_{\rm r}^2}\right)^2\right]~, \label{c12scr}
\end{eqnarray}
where for lattices in consideration
\begin{equation}
\chi_{\rm bcc}=0.00661835~, \qquad  \chi_{\rm fcc}=0.00952744~.
\end{equation}
Eqs. (\ref{Pscr}-\ref{c12scr}) are valid for any lattice with cubic symmetry. 
For the bcc lattice these results were obtained in Ref.~\onlinecite{B15} and agrees with ours. As expected, the coefficients at $b^2$ and $b^4$ depend on  the polarization correction to the electrostatic energy $\eta$.

Our results for $c_{11}$ and $c_{12}$ differ from those presented in Ref.~\onlinecite{RKG88}.
It is due to the fact that this parameters depend on the volume change and, consequently, on the background properties. While for deformations with the conservation of volume, it does not matter from what the background consists of and what particles form a crystal lattice.

For the stretching the volume stays constant at $c_2=1/c_1$. It allows to compute $c_{11}-c_{12}$, which can be considered as a function of $\kappa a$. The values of $c_{11}-c_{12}$ for several $\kappa a$ are presented in Table \ref{Tab1}.
\begin{table}
\caption{\label{Tab1} Values of $c_{11}-c_{12}$ and $c_{44}$ in units of $nZ^2e^2/a$ for bcc and fcc lattices.}
\begin{ruledtabular}
\begin{tabular}{ccccc}
&  \multicolumn{2}{c}{bcc lattice}  & \multicolumn{2}{c}{fcc lattice} \\          
\hline
$\kappa a$ & $c_{11}-c_{12}$ & $c_{44}$ & $c_{11}-c_{12}$ & $c_{44}$  \\
\hline
0.0 & 0.0489772 & 0.1827696 & 0.0413464 & 0.185301 \\
0.5 & 0.0454427 & 0.172799 & 0.0398645 & 0.17464 \\
1.0 & 0.0363862  & 0.146462 & 0.0356839 & 0.146677 \\
1.5 & 0.0253054 & 0.112199 & 0.0295833 & 0.110775 \\
2.0 & 0.0154092 & 0.0785928 & 0.0226992 & 0.0761958 \\
2.5 & 0.00826454  & 0.0510008 & 0.0161631 & 0.0484181 \\
3.0 & 0.00389811 & 0.0310557 & 0.0107426 & 0.0288187 \\
3.5 & 0.00158472 & 0.0179504 & 0.00671567 & 0.0162653 \\
4.0 & 0.000515049  & 0.00994519 & 0.00398121 & 0.0087949 \\
4.5 & 0.0000910451 & 0.00532368 & 0.00225578 & 0.00459394 \\
\end{tabular}
\end{ruledtabular}
\end{table}

The difference between data presented in Table III in Ref.~\onlinecite{RKG88} and our calculations is less than $0.5\%$ for $\kappa a < 4.4$. All quantities in Ref.~\onlinecite{RKG88} are presented in $n^{-1/3}$ units, so their $\lambda=7$ corresponds to our $\kappa a \approx 4.34$. The bcc lattice is stable at $\kappa a < 4.76$, and $c_{11}-c_{12}$ becomes negative for higher $\kappa a$.  

For $\kappa a \lesssim 3$ the parameter $c_{11}-c_{12}$ could be fitted as
\begin{equation}
c_{11}-c_{12}= n \frac{Z^2e^2}{a}\frac{k_0 + k_1 (\kappa a)^2 + k_2 (\kappa a)^4}{1 + k_3 (\kappa a)^2 + k_4 (\kappa a)^4}~, 
\label{fit}
\end{equation}
where the fitting parameters $k_{0} \dots k_{4}$ are presented in Table \ref{Tab2}.
Approximation was carried out over more than 30 points. The maximum error is less than $0.5\%$ for the bcc lattice and $0.02\%$ for the fcc lattice. The fcc lattice is stable even at $\kappa a =15$, so its energy dependence on $\kappa a$ is more smooth.
\begin{table}
\caption{\label{Tab2}Fitting parameters $k_{0 \dots 4}$ for bcc and fcc lattices.}
\begin{ruledtabular}
\begin{tabular}{ccccc}
&  \multicolumn{2}{c}{bcc lattice}  & \multicolumn{2}{c}{fcc lattice} \\          
\hline
$k_i$ & $c_{11}-c_{12}$ & $c_{44}$ & $c_{11}-c_{12}$ & $c_{44}$  \\
\hline
$k_0$ & 0.0489772 & 0.18277 & 0.0413464 & 0.185301  \\
$k_1$ & $-0.00510682$ & $-0.0113456$ & $-0.00204533$ & $-0.0126294$ \\
$k_2$ & 0.000149185 & 0.000226914 & 0.000029972 & 0.000281284 \\
$k_3$ & 0.196328 & 0.162907 & 0.0960401 & 0.169997 \\
$k_4$ & 0.0138269 & 0.00892591 & 0.00613073 & 0.00923629 \\
\end{tabular}
\end{ruledtabular}
\end{table}

\begin{figure}
    \center{\includegraphics[width=\linewidth]{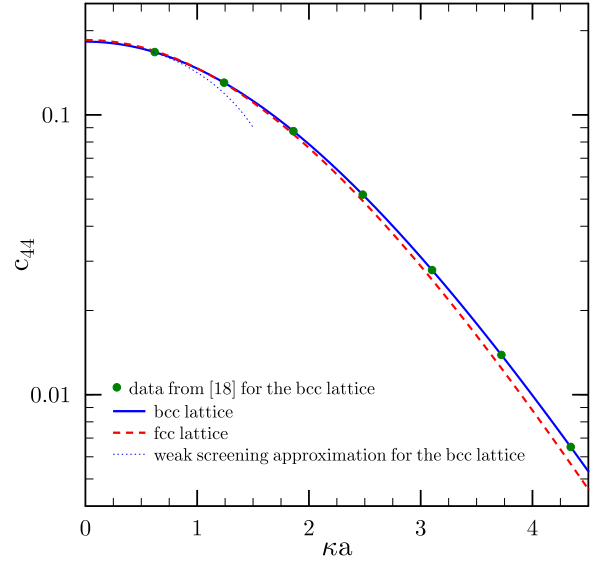}}
    \caption{The elastic modulus $c_{44}$ in units of $nZ^2e^2/a$ for bcc (solid blue line) and fcc (red line) lattices obtained from Eq. (\ref{UTF}), see details in text. Thin dotted line shows the weak screening approximation for the bcc lattice, see Eq. (\ref{fitb}). Data from Ref.~\onlinecite{RKG88} is plotted by dots.}
    \label{fig:1}
\end{figure}
Another type of deformation which conserves the volume of an elementary cell is the shift. Under this deformation the vector $a_{\rm l}(l_1,l_2,l_3)$ translates as
\begin{equation}
a_{\rm l}(l_1,l_2,l_3)\to a_{\rm l}\left(l_1+c_x l_3, l_2,  l_3,\right)~.
\end{equation}
where $c_x$ is the parameter of deformation.
Using this deformation, we can obtain the third elastic modulus:
\begin{equation}
c_{44}= n\frac{Z^2e^2}{a} s^{xyxy}~,
\end{equation}
where $s^{xyxy}$ could be obtained from the quadratic term of the expansion of the electrostatic energy into $c_x$. 

$c_{44}$ is plotted in Fig. \ref{fig:1} in units of $nZ^2e^2/a$ for $\kappa a < 4.5$, and for several values of $\kappa a$ is presented in Tab. \ref{Tab1}. One can see that the difference between results for different lattices is small.

At $\kappa a \ll 1$ equation for $c_{44}$ can be written as 
\begin{eqnarray}
c_{44, \rm bcc}&=&n\frac{Z^2e^2}{a}\left(0.1827696-0.041120(\kappa a)^2\right)~, \label{fitb} \\
c_{44, \rm fcc}&=&n\frac{Z^2e^2}{a}\left(0.185301-0.04413(\kappa a)^2\right)~.
\end{eqnarray}
The same equation for the bcc lattice was obtained in Ref.~\onlinecite{B15}.
Eq. (\ref{fitb}) is plotted in Fig. \ref{fig:1} by thin line. At $\kappa a = 0.75$ this approximation differs from the exact value by only $1.12\%$, but at higher $\kappa a$ differences increase significantly.

For $\kappa a \lesssim 3$ the coefficient $c_{44}$ can be fitted by a similar equation as for $c_{11}-c_{12}$ [Eq. (\ref{fit})]. Fitting parameters are presented in Table \ref{Tab2}. The maximum deviation from the precise result does not exceed $0.1 \%$. 

At $\kappa a =0$ our calculations for $c_{44}$ agree with previous results Refs.~\onlinecite{F36,B11,K19}. At $\kappa a >0$ they agree with data obtained in Ref.~\onlinecite{RKG88}. The maximum deviation does not exceed one percent for any $\kappa a$ in consideration. It could be seen in Fig. \ref{fig:1}, where data obtained in Ref.~\onlinecite{RKG88} is shown by green dots. In the selected scale the difference from our results is not visible.

Parameter $\Delta/c_{44}\equiv 2-(c_{11}-c_{12})/c_{44}$ also coincides with Ref.~\onlinecite{RKG88}. For the bcc lattice the ratio $\Delta/c_{44}$ increases with $\kappa a$ from 1.732, while for the fcc lattice this ratio decreases from 1.777 and reaches 1.587 at $\kappa a = 3.5$. The experimental results obtained in Ref.~\onlinecite{RSW15} for colloidal Wigner crystals give $\Delta/c_{44} \lesssim 1.5$ and show a decrease with $\kappa a$ at $\kappa a \approx 3-4$, which agrees with our theory qualitatively. For other theoretical and experimental approaches Ref.~\onlinecite{L21} the typical value for  $\Delta/c_{44}$ lies around $0.7-1.0$.

In addition we can calculate the effective shear modulus determined as a Voigt average (e.g., Ref.~\onlinecite{B11})
\begin{equation}
\mu_{\rm eff}\equiv \frac{1}{5}(c_{11}-c_{12}+3c_{44})~.
\end{equation}
This definition seems to be most applicable to degenerate stars and in good agreement with the linear mixing rule\cite{K19}.

For bcc and fcc lattices with $\kappa a=0$ it was obtained in Refs.~\onlinecite{B11,K19} that 
\begin{equation}
\mu_{\rm eff}=-\frac{2}{15} n\frac{Z^2e^2}{a} \zeta~.
\label{mu}
\end{equation}
Recently validity of this equality for any isotropic lattice was analytically proved in Ref.~\onlinecite{C21}.

The first correction to Eq. (\ref{mu}) for the weak screening could be written as
\begin{equation}
\mu_{\rm eff}^{\rm scr}=\frac{4}{15}n\frac{Z^2e^2}{a} \eta(\kappa a)^2~.
\label{mus}
\end{equation}
For the bcc lattice it was obtained numerically in Ref.~\onlinecite{B15}.  Eq. (\ref{mus}) is apparently fair for any isotropic lattice\cite{C22}. 

At $\kappa a \ll 1$ for the degenerate electron background in the ultrarelativistic limit the effective shear modulus of the bcc lattice can be written as 
\begin{equation}
\mu_{\rm eff}=0.119457 n\frac{Z^2e^2}{a}(1-0.007924 Z^{2/3})~.
\label{ult}
\end{equation}
As mentioned earlier in Ref.~\onlinecite{B15}, this polarization correction is $20\%$ smaller than the value obtained in Ref.~\onlinecite{KP13}. Such large difference could be important for studies of neutron-star crust oscillations, alas in some papers (e.g., Ref.~\onlinecite{T17,T21}) not precise enough expressions are still used.

For higher $\kappa a$ the effective shear modulus could be obtained from our approximations and  parameters listed in Table \ref{Tab2}.

The effective shear modulus for the bcc lattice has been studied in Ref.~\onlinecite{HH08} by molecular dynamics simulations. At $\kappa a \approx 0.5705$ and $T=0$ the authors found that the effective shear modulus is $0.1108nZ^2e^2/a$. Our calculations give $\mu_{\rm eff} \approx 0.1105nZ^2e^2/a$, so the agreement is very good. 

It should be mentioned that in Ref.~\onlinecite{Kh19} it was shown that sound velocities and, respectively, elastic parameters of dusty Yukawa systems are related to the electrostatic energy and its derivatives with respect to kappa. For instance, according to Ref.~\onlinecite{Kh19} the effective shear modulus is equal
\begin{equation}
\mu_{\rm eff}=\frac{1}{15}n\frac{Z^2e^2}{a}\left[\kappa^2 \frac{\partial^2 M}{\partial \kappa^2}+2 \kappa \frac{\partial M}{\partial \kappa}-2M\right]~,
\label{ultK}
\end{equation}
where $M$ is the Madelung constant ($M \approx \zeta+\eta (\kappa a)^2$ at $\kappa a \ll 1$). Then for all $\kappa a$ in consideration if we substitute into Eq. (\ref{ultK}) Madelung constant calculated via Eq. (\ref{UTF}), we will get exactly the same values (at used calculation accuracy of 6 significant digits) for $\mu_{\rm eff}$ that our method gives. Despite of the significant difference in approaches to the calculation, the obtained agreement confirms the correctness of our method.

\section{Conclusions}
In the current paper we discussed elastic properties of Yukawa crystals. For the first time they were calculated analytically for systems with arbitrary screening parameter. 

For crystals with the degenerate background our method can be applied only for the first order polarization correction $\propto(\kappa_{\rm TF} a)^2$ to the uniform background case. For the bcc lattice results are well consistent with previous calculations performed in Ref.~\onlinecite{B15} and improve them, while for the fcc lattice the present results are new. For both lattices the effective shear modulus in this approximation is equal 
\begin{equation}
\mu_{\rm eff}=\left(-\frac{2}{15} \zeta+\frac{4}{15} \eta(\kappa a)^2 \right)n\frac{Z^2e^2}{a}~,.
\end{equation}
where  $\zeta$ is a Madelung constant, $\eta$ is a polarization correction to this Madelung constant.

For the case when the neutralizing background is degenerated we have calculated constants $c_{11}-c_{12}$ and $c_{44}$ and obtained convenient approximations for them at $\kappa a \lesssim 3$. These two quantities for bcc and fcc lattices are in a good agreement with results of Ref.~\onlinecite{RKG88}, which allows us to conclude that the elastic moduli of dusty Yukawa crystals can be successfully calculated not only using molecular-dynamic simulation, but also analytically.

\begin{acknowledgments}
The author is deeply grateful to A.~Y. Potekhin and A.~I. Chugunov for help and discussions, Ksenia K. for inspiration, and to an anonymous referee for valuable comments.

This research was supported by The Ministry of Science and Higher
Education of the Russian Federation (Agreement with Joint Institute for
High Temperatures RAS No.\,075-15-2020-785 dated September 23, 2020).

\end{acknowledgments}

\end{document}